\documentclass[12pt]{article}

\usepackage{amsmath,amsfonts,amssymb}

\def\ua{\underline{a}}
\def\ub{\underline{b}}

\def\bB{\mathbf{B}}

\def\be{\begin{equation}}

\def\ee{\end{equation}}

\def\bea{\begin{eqnarray}}

\def\eea{\end{eqnarray}}

\def\tmH{\tilde{\mH}}

\def\mH{\mathcal{H}}

\newcommand{\mK}{\mathcal{K}}

\newcommand{\mG}{\mathcal{G}}

\newcommand{\mL}{\mathcal{L}}

\def\pb #1{\left\{#1\right\}}

\begin{document}

	\begin{titlepage}

		\vskip 0.4 cm
		
		\begin{center}
			{\Large{ \bf New Non-Relativistic String in $AdS_5\times S^5$}
			}
			
			\vspace{1em}  
			
			\vspace{1em} J. Kluso\v{n} 			
			\footnote{Email addresses:
			 klu@physics.muni.cz  }\\
			\vspace{1em}
\textit{Department of Theoretical Physics and
				Astrophysics, Faculty of Science,\\
				Masaryk University, Kotl\'a\v{r}sk\'a 2, 611 37, Brno, Czech Republic}
			
			\vskip 0.8cm
			
%
%
%
%
			%
			%
			
			\vskip 0.8cm
			
		\end{center}

		\begin{abstract}
We study non-relativistic limit of $AdS_5\times S^5$ background and determine
corresponding Newton-Cartan fields. We also find canonical form of the new
non-relativistic string in this background and discuss its formulation in the uniform light-cone gauge.	

		\end{abstract}
		
		\bigskip
		
	\end{titlepage}
	
	\newpage

\section{Introduction and Summary}
In the past few years it is observed renewed interest in the study of non-relativistic string theories in Newton-Cartan (NC) formulation
\cite{Cartan:1923zea,Cartan:1924yea}. Basically,  NC gravity provides
covariant description of Newton's law. However it is very remarkable
that NC description can be extended also into more broader class of theories, 
as for example field theory and string theory. In fact, non-relativistic string theory 
was originally introduced in 2000 in two  papers
 \cite{Gomis:2000bd,Danielsson:2000gi}. These theories were defined without
 Newton-Cartan formalism which was firstly introduced in the context of string theory in the 
 remarkable paper \cite{Andringa:2012uz}, for related works, see
 for example 
\cite{Bidussi:2021ujm,Bergshoeff:2021tfn,Hartong:2021ekg,Blair:2021ycc,Bergshoeff:2021bmc,Gomis:2020izd,
	Kluson:2020rij,Yan:2019xsf,Bergshoeff:2019pij,Harmark:2019upf,Gallegos:2019icg,
	Gomis:2019zyu,Kluson:2019uza,Kluson:2018vfd,Bergshoeff:2018vfn,Kluson:2018grx,Bergshoeff:2018yvt}.  Different formulation of non-relativistic string 
was presented in  
\cite{Harmark:2017rpg} that was based on T-duality of string theory 
in the background with null isometry
\footnote{For  related works, see
\cite{Kluson:2021djs,Kluson:2021pux,Fontanella:2021hcb,Kluson:2021sym,Harmark:2020vll,Kluson:2020aoq,Hansen:2020pqs,Kluson:2019xuo,Kluson:2019avy,Hansen:2019pkl,Kluson:2018egd,Harmark:2019upf}.}. In \cite{Bidussi:2021ujm} new form of non-relativistic
string theory was proposed that has interesting property that non-relativistic string
naturally couples to two form field $m_{MN}$ in the similar way as  non-relativistic particle couples to mass form $m_M$. 

This new proposal of non-relativistic string was further studied in \cite{Kluson:2021qqv} where canonical formulation of this theory was found. We also analysed the possibility to impose uniform light-cone gauge  on this theory and found Hamiltonian on reduced phase space. Uniform light-cone 
gauge was used in
 \cite{Frolov:2019nrr,Arutyunov:2006gs,Frolov:2006cc,Arutyunov:2005hd}
  \footnote{For review, see for example \cite{Arutyunov:2009ga}.} where it was shown that it is very efficient for the study of dynamics of the relativistic
  string in $AdS_5\times S^5$. We showed in \cite{Kluson:2021qqv}
  that such an uniform light cone gauge fixing can be imposed in case of non-relativistic string as well at least at the formal level. 
  
 In this paper we continue the analysis of new non-relativistic string when
 we focus on its explicit formulation as non-relativistic limit of $AdS_5\times S^5$ background. Our starting point is an important paper \cite{Gomis:2005pg}
  where non-relativistic strings in $AdS_5\times S^5$ was defined by specific limiting procedure. We combine this procedure with the definition of Newton-Cartan fields as was given in \cite{Bidussi:2021ujm} and we will be able
  to find Newton-Cartan background fields for non-relativistic limit of $AdS_5\times S^5$. Explicitly, we find $2\times 2$ twobein $\tau_M^{ \ A}$ together with field $\pi_M^{ \ A}$ that was introduced in 
  \cite{Bidussi:2021ujm}. Then we will be able to determine two form $m_{MN}$ and hence corresponding Hamiltonian. As the next step we study gauge fixed form of the theory when we impose uniform light-cone gauge. Solving Hamiltonian constraint we determine Hamiltonian on the reduced phase space. We find that the structure of this Hamiltonian depends on the 
  free parameter that defines generalized uniform light cone gauge \cite{Frolov:2019nrr,Arutyunov:2006gs,Frolov:2006cc,Arutyunov:2005hd}. Then we study equations of motion on the reduced phase space. We show that it is possible to have configuration with all free fields to be equal to zero and concentrate on the dynamics of the mode $z_1$. However then we find that  the equation of motion for $z_1$ is solved by arbitrary function and hence does not determine dynamics of $z_1$ at all. We mean that this is a sign that the present form of  uniform light cone gauge  is not suitable for the specific form of non-relativistic string studied in this paper. 
  
  Then in order to study properties of non-relativistic string in more details we focus on its Lagrangian equations of motion. We determine their form for general background. Since these equations of motion are rather complicated in the full generality we restrict ourselves to  an analysis of the dynamics of single coordinate $z_1$. We find solution that has formally  the same form as solution found recently in \cite{Fontanella:2021btt} however there is a crucial difference since we consider extended string along non-compact coordinate and hence we should interpret this solution as the string with infinite number of spikes.
  
  Let us outline our results. The main goal of this paper was to find Hamiltonian for new non-relativistic string in  $AdS_5\times S^5$ background. Performing appropriate limit we determined components of Newton-Cartan fields and then we obtained corresponding Hamiltonian. We studied its gauge fixed form and we argued
  that the uniform light cone gauge could be too restrictive to obtain interesting
  dynamics. On the other hand it is possible that different gauge fixing procedure, as for example static gauge, could lead to non-trivial dynamics of free world-sheet fields on $AdS_2$ background as was shown in \cite{Gomis:2005pg}. We also studied Lagrangian equations of motion and we found solution corresponding to string extended along one free spatial coordinate and non-trivial dynamics along of $z_1$ coordinate that agrees with solution found in \cite{Fontanella:2021btt}. 
  
  This paper is organized as follows. In the next section (\ref{second}) we review basics facts about new non-relativistic string and its canonical formulation. Then in section (\ref{third}) we find its form in non-relativistic limit of $AdS_5\times S^5$. In section (\ref{fourth}) we study properties of this string in uniform light-cone gauge. In section (\ref{fifth}) we determine Lagrangian equations of motion and study correspoding solution. Finally in appendix (\ref{sixth}) we perform non-relativistic limit in coordinates that were used \cite{Gomis:2005pg} and find corresponding Hamiltonian.

\section{Review of New Non-Relativistic String and Its Canonical Formulation}\label{second}
In this section we review basic facts about new non-relativistic string
action as was proposed in \cite{Bidussi:2021ujm} and that has the form
\begin{equation}
S=-\frac{T}{2}\int d^2\sigma \sqrt{-\tau}[
\tau^{\alpha\beta}h_{MN}+\epsilon^{\alpha\beta}m_{MN}]\partial_\alpha x^M
\partial_\beta x^N \ . 
\end{equation}
We firstly describe derivation of this action, following  \cite{Bidussi:2021ujm}. Let us introduce relativistic vielbein $e_M^{ \ \underline{a}}$ so that 
 target space metric has the form
\begin{equation}\label{gmunu}
g_{MN}=e_M^{ \ \ua}e_N^{ \ \ub}\eta_{\ua\ub}  \ , 
\end{equation}
where we use the similar notation as in  \cite{Bidussi:2021ujm} so that  frame indices are $\ua,\ub=0,\dots,9$ and where $\eta_{\ua\ub}=
\mathrm{diag}(-1,1,\dots,1)$. Note that space-time indices are $M,N=0,1,\dots,9$. Following 
\cite{Bidussi:2021ujm} we also introduce parametrization 
of NSNS two form $B_{MN}$ as
\begin{equation}
B_{MN}=\frac{1}{2}\eta_{\ua\ub}(e_M^{ \ \ua}\pi_N^{ \ \ub}-
e_N^{ \ \ua}\pi_M^{ \ \ub}) \ . 
\end{equation}
To begin with let us write
 Nambu-Goto form of the action for relativistic string in general background
\begin{equation}
S=-cT_F\int d^2\sigma \sqrt{-\det g_{\alpha\beta}}-cT_F
\int d^2\sigma \frac{1}{2}\epsilon^{\alpha\beta}B_{\alpha\beta} \ , 
\end{equation}
where $g_{\alpha\beta}=g_{MN}\partial_\alpha x^M\partial_\beta x^N \ , \quad B_{\alpha\beta}=B_{MN}\partial_\alpha x^M
\partial_\beta x^N$ and where $\epsilon^{01}=1=-\epsilon_{01}$ and $T_F$ is string tension.
 As in 
\cite{Bidussi:2021ujm} we introduce
indices $\ua=(A,a)$ corresponding to directions longitudinal and transverse to 
string world-sheet where $A=0,1$ are longitudinal and $a=2,\dots,9$ are transverse. 
Then we have
\begin{equation}\label{eMua}
e_M^{ \ \ua}=(cE_M^{ \ A},e_M^{ \ a}) \ , \quad 
\pi_M^{ \ \ua}=(c\Pi_M^{ \ A},\pi_M^{ \ a}) \ 
\end{equation}
so that
\begin{eqnarray}\label{gBexp}
& &g_{\alpha\beta}=c^2 \eta_{AB}E_\alpha^{ \ A}E_\beta^{ \ B}+
\delta_{ab}e_\alpha^{ \ a}e_\beta^{ \ b}  \ ,  \nonumber \\
& &B_{\alpha\beta}=\frac{1}{2}c^2\eta_{AB}(E_\alpha^{ \ A}\Pi_\beta^{ \ B}-
E_\beta^{ \ A}\Pi_\alpha^{ \ B})+\frac{1}{2}\delta_{ab}(e_\alpha^{ \ a}
\pi_\beta^{ \ b}-e_\beta^{ \ a}\pi_\alpha^{ \ b}) \ . \nonumber \\
\end{eqnarray}
We further 
 parametrize longitudinal components in the following way
\begin{eqnarray}\label{EmuAi}
E_M^{ \ A}=\tau_M^{ \ A}+\frac{1}{2c^2}\pi_M^{ \ B}\epsilon_B^{ \ A} \ , 
\quad
\Pi_M^{ \ A}=\epsilon^A_{ \ B}\tau_M^{ \ B}+\frac{1}{2c^2}\pi_M^{ \ A} \ , 
\nonumber \\
\end{eqnarray}
where $\epsilon_B^{\ A}=\epsilon_{BC}\eta^{CA}$.
 Inserting (\ref{EmuAi}) into (\ref{gmunu})  we finally get
\begin{eqnarray}
& &g_{\alpha\beta}=c^2 \tau_{\alpha\beta}
+\frac{1}{2}\eta_{AB}(\tau_\alpha^{ \ A}\pi_\beta^{ \ C}\epsilon_{C}^{ \ B}+
\tau_\beta^{ \ B}\pi_\alpha^{ \ C}\epsilon_{C}^{ \ B})+h_{\alpha\beta}
\nonumber \\
& &+\frac{1}{4c^2}\eta_{AB}\pi_\alpha^{ \ C}\epsilon_C^{ \ A}\pi_\beta^{  \ D}
\epsilon_D^{ \ B} \ , \nonumber \\
\end{eqnarray}
where
\begin{equation}
\tau_{\alpha\beta}=\tau_\alpha^{\  A}
\tau_\beta^{ \ B}\eta_{AB} \ , \quad 
h_{\alpha\beta}=e_\alpha^{ \ a}e_\beta^{ \ b}\delta_{ab} \ . 
\end{equation}
Then it can be shown that the resulting non-relativistic action has the form
\begin{equation}\label{Snonfinal}
S=-\frac{T}{2}\int d^2\sigma [\sqrt{-\tau}\tau^{\alpha\beta}h_{\alpha\beta}+
\epsilon^{\alpha\beta}m_{\alpha\beta}]  \ , 
\end{equation}
where we introduced rescaled tension  $cT_{F}=T$ and we have taken the limit $c\rightarrow \infty$. Finally we also introduced matrix $\tau^{\alpha\beta}=\tau^\alpha_{ \ A}\tau^\beta_{ \ B}\eta^{AB}$ which is $2\times 2$ matrix  inverse
to $\tau_{\alpha\beta}$. We also introduced $2\times 2$ twobein $\tau^\alpha_{ \ A}$ that obeys the condition 
\begin{equation}
\tau_\alpha^{ \ A}\tau^\beta_{ \ A}=\delta_\alpha^\beta \ , 
\quad
\tau_\alpha^{ \ A}\tau^\alpha_{ \ B}=\delta^A_B \ . 
\end{equation}
Note that  $m_{\alpha\beta}=m_{MN}\partial_\alpha x^M\partial_\beta x^N$ that is written in (\ref{Snonfinal}) is defined as
\begin{equation}
m_{MN}=\frac{1}{2}\eta_{AB}[\tau_M^{ \ A}\pi_N^{ \ B}-
\tau_N^{ \ A}\pi_M^{ \ B}]+
\frac{1}{2}\delta_{ab}[e_M^{ \ a}\pi_N^{ \ b}-e_N^{ \ a}\pi_M^{ \ b}] \ .
\end{equation}
The canonical form of the action (\ref{Snonfinal}) was recently analysed in \cite{Kluson:2021qqv} where it was shown that the Hamiltonian is sum of two first class constraints 
\begin{equation}
H=\int d\sigma (N^\tau \mH_\tau+N^\sigma \mH_\sigma) \ , 
\end{equation}
where 
\begin{eqnarray}
& &\mH_\sigma=p_M\partial_\sigma x^M \approx 0 \  , \nonumber \\
& &\mH_\tau=-2T\Pi_M \tau^M_{ \ A}\eta^{AB}\epsilon_{BD}\tau_\sigma^{ \ D}+T^2h_{\sigma\sigma}+\Pi_M h^{MN}\Pi_N\approx 0 \ , \nonumber \\
\end{eqnarray}
where $\Pi_M$ is defined as
\begin{equation}
\Pi_M=p_M+Tm_{MN}\partial_\sigma x^N \ . 
\end{equation}
After the review of main properties of new non-relativistic string action 
we proceed to its explicit form when we consider  non-relativistic
limit  of $AdS_5\times S^5$.
\section{Non-Relativistic $AdS\times S^5$ Background}\label{third}
Following general prescription reviewed in the previous section we would like
to find Newton-Cartan fields for non-relativistic limit of $AdS_5\times S^5$. 
Let us now consider $AdS_5\times S^5$ background in Cartesian global coordinates where line element has the form
\begin{equation}
ds^2=g_{TT}dT^2+g_{Z_iZ_j}dZ^idZ^j+g_{\Phi\Phi}d\Phi^2+g_{Y_iY_j}dY^idY^j \ , 
\end{equation}
where 
\begin{eqnarray}
& &g_{TT}=-\left(\frac{1+\frac{Z^2}{4R^2}}{1-
\frac{Z^2}{4R^2}}\right)^2 \ , \quad Z^2\equiv Z_i Z_i \ ,  \quad 
g_{Z_iZ_j}=\frac{1}{\left(1-\frac{Z^2}{4R^2}\right)^2}\delta_{ij} \ , 
\nonumber \\
& &g_{\Phi\Phi}=\left(\frac{1-\frac{Y^2}{4R^2}}{1+\frac{Y^2}{4R^2}}\right)^2 \ , 
\quad g_{Y_iY_j}=\left(\frac{1}{1+\frac{Y^2}{4R^2}}\right)^2\delta_{ij} \ , \quad
Y^2\equiv Y_i Y^i \ , \nonumber \\
\end{eqnarray}
where $i,j=1,2,3,4$ and 
where $R$ is common radius of $AdS_5$ and $S^5$.
It is convenient to write this line element as $ds^2=e^{ \ \bar{a}}e^{ \ \bar{b}}\eta_{\bar{a} \bar{b}}$. Then, following \cite{Gomis:2005pg}, we define non-relativistic limit as
\begin{equation}\label{rescal}
T\rightarrow c t \ , \quad Z_1\rightarrow c z_1 \ , \quad Z_m\rightarrow z_m
,\quad \Phi\rightarrow \phi  \ , 
\quad Y_i\rightarrow y_i \ , \quad  R=cR_0 \ , 
\end{equation}
where $m=2,3,4$ and where non-relativistic limit corresponds to  $c\rightarrow \infty$. We start with the vielbein 
$e^{ \ 0}$ that after rescaling (\ref{rescal}) takes the form 
\begin{eqnarray}
& &e^{ \ 0}=\frac{1+\frac{Z^2}{4R^2}}{1-\frac{Z^2}{4R^2}}dX^0
=c\frac{1+\frac{z_1^2}{4R_0^2}}{1-\frac{z_1^2}{4R_0^2}}dt
+\frac{1}{c}\frac{z_m z_m}{2R_0^2(1-\frac{z_1^2}{4R_0^2})^2}dt \ . 
\nonumber \\
\end{eqnarray}
Then comparing this expression with (\ref{eMua}) and (\ref{EmuAi}) we can 
identify 
\begin{eqnarray}\label{taut0}
\tau_t^{ \ 0}=
\frac{1+\frac{z_1^2}{4R_0^2}}{1-\frac{z_1^2}{4R_0^2}} \ , \quad 
\pi_t^{ \ 1}=-\frac{z_m z_m}{R_0^2(1-\frac{z_1^2}{4R_0^2})^2} \ . 
\nonumber \\
\end{eqnarray}
In the same way we proceed with $e^{ \ 1}$
\begin{eqnarray}
e^{ \ 1}=\frac{1}{\left(1-\frac{Z^2}{4R^2}\right)}dZ^1=
c\frac{1}{1-\frac{z_1^2}{4R_0^2}}dz_1+\frac{1}{c}\frac{1}{4R_0^2}\frac{z_m z_m}{
(1-\frac{z_1^2}{4R_0^2})^2}dz_1 \nonumber \\
\end{eqnarray}
so that comparing with (\ref{EmuAi}) we get
\begin{equation}\label{tauz11}
\tau_{z_1}^{ \ 1}=
\frac{1}{1-\frac{z_1^2}{4R_0^2}}  \ , \quad \pi_{z_1}^{ \ 0}=-
\frac{z_m z_m}{
	2R_0^2(1-\frac{z_1^2}{4R_0^2})^2} \ . 
\end{equation}
In case of the $e^{ \ m}$ the situation is simpler 
\begin{eqnarray}
e^m=\frac{1}{1-\frac{Z^2}{4R^2}}dZ_m=
\frac{1}{1-\frac{z_1^2}{4R_0^2}}dz_m
\nonumber \\
\end{eqnarray}
and hence
\begin{equation}
e_{z_m}^{ \ n}=\frac{1}{1-\frac{z_1^2}{4R_0^2}}\delta_m^{ \ n} \ . 
\end{equation}
In the same way we obtain
\begin{equation}
e_{\phi}^{ \ \Phi}=1 \ , \quad e_{y_i}^{ \ j}=\delta_i^{ \  j} \ . 
\end{equation}
It is important to stress that due to the fact that $R=cR_0
\rightarrow \infty$ the $\phi$ coordinate is effectively non-compact since 
original variable $\Phi$ was periodic with period $2\pi R$. 

Now we are ready to proceed to find corresponding Hamiltonian. We firstly determine
components of $m_{MN}$ that, using
(\ref{taut0}) and (\ref{tauz11}) have following non-zero elements
\begin{eqnarray}
m_{tz_1}=
\frac{(1+\frac{z_1^2}{4R_0^2})z_m z_m}{4R_0^2(1-\frac{z_1^2}{4R_0^2})^3} \ . 
\nonumber \\
\end{eqnarray}
 Further, from the relation 
$\tau_M^{ \ A}\tau^M_{ \ B}=\delta^A_B$ we obtain that there are non-zero components
of matrix inverse $\tau^M_{ \ A}$ equal to 
\begin{eqnarray}
\tau^t_{ \ 0}=\frac{1}{\tau_t^{ \ 0}}=
\frac{1-\frac{z_1^2}{4R_0^2}}{1+\frac{z_1^2}{4R_0^2}}  \ , \quad 
\tau^{z_1}_{ \ 1}=
	1-\frac{z_1^2}{4R_0^2} \ .  \nonumber \\
\end{eqnarray}
Taking all these results into account we obtain that the  Hamiltonian constraint of non-relativistic string is equal to 
\begin{eqnarray}
& &\mH_\tau=-2T \Pi_t\tau^t_{ \ 0}\tau_{z_1}^{ \ 1}\partial_\sigma z_1-2T\Pi_{z_1}
\tau^{z_1}_{ \ 1}\tau_t^{ \ 0}\partial_\sigma t+p_{z_m}(1-\frac{z_1^2}{4R_0^2})^2
p_{z_m}+T^2\partial_\sigma z^m 
\frac{1}{(1-\frac{z_1^2}{4R_0^2})^2}\partial_\sigma z^m+
\nonumber \\
& &+p_{\phi}^2+(p_{y_1})^2+T^2(\partial_\sigma \phi)^2+T^2(\partial_\sigma y_i)^2 \ ,  \nonumber \\
\end{eqnarray}
where
\begin{equation}
\Pi_t=p_t+Tm_{tz_1}\partial_1 z_1 \ , \quad 
\Pi_{z_1}=p_{z_1}+Tm_{z_1 t}\partial_1 t \ . 
\end{equation}
%
%
As the next step we would like to find uniform light-cone gauge fixing form of non-relativistic string. Discussion of the general case 
was performed in \cite{Kluson:2021qqv} and here we focus on the non-relativistic
limit of $AdS_5\times S^5$. In order to impose uniform light-cone gauge the
background should possesses two abelian isometries where one of them is $t$. From the form of the non-relativistic background  it is clear that the second one can be  either $\phi$ or $y_i$ where now $\phi$ is non-compact. Without lost of generality we select $\phi$ as the second coordinate with isometry.

As the next step we introduce light-cone coordinates and momenta \cite{Frolov:2019nrr,Arutyunov:2009ga,Arutyunov:2006gs}
\begin{eqnarray}
& &x^-=\phi-t \ , \quad x^+=\frac{1}{2}(\phi+t)+\alpha x^- \ , 
\nonumber \\
& &p_+=p_\phi+p_t \ , \quad p_-=\frac{1}{2}(p_\phi-p_t)-\alpha p_+ \ , \nonumber \\
\end{eqnarray}
with inverse relations
\begin{eqnarray}
& &\phi=x^++x^-(\frac{1}{2}-\alpha) \ , \quad t=x^+-x^-(\frac{1}{2}+\alpha) \ ,  \nonumber \\
& & p_t=p_+(\frac{1}{2}-\alpha)-p_- \ , \quad p_\phi=p_-+p_+(\frac{1}{2}+\alpha) \ ,
\nonumber \\
\end{eqnarray}
where $\alpha$ is free parameter.
Let us insert these relations to the Hamiltonian constraint given above and we get
\begin{eqnarray}\label{mHtaulc}
& &\mH_\tau=-2T((p_+(\frac{1}{2}-\alpha)-p_-)+Tm_{tz_1}\partial_\sigma z_1)
\tau^t_{ \ 0}\tau_{z_1}^{ \ 1}\partial_\sigma z_1-\nonumber \\
&&-2
T(p_{z_1}+Tm_{z_1t}(\partial_\sigma x^+-\partial_\sigma x^-(\frac{1}{2}+\alpha)))
\tau^{z_1}_{ \ 1}\tau_t^{ \ 0}\partial_\sigma (x^+-x^-(\frac{1}{2}+\alpha))+\nonumber \\
& &+p_{z_m}(1-\frac{z_1^2}{4R_0^2})^2
p_{z_m}+T^2\partial_\sigma z^m 
\frac{1}{(1-\frac{z_1^2}{4R_0^2})^2}\partial_\sigma z^m+
\nonumber \\
& &+(p_{y_1})^2+T^2(\partial_\sigma y_i)^2
+(p_-+p_+(\frac{1}{2}+\alpha))^2+T^2(\partial_\sigma x^++\partial_\sigma x^-(\frac{1}{2}-\alpha) )^2 \ . 
\nonumber \\
\end{eqnarray}
Now we are ready to impose uniform light-cone gauge by introducing following
gauge fixing functions \cite{Frolov:2019nrr}
\begin{equation}\label{gff}
\mG^+\equiv x^+-\tau\approx 0 \ , \quad  \mG^-=p_--T\approx 0 \ , \quad 
a=\frac{1}{2}+\alpha \ .
\end{equation}
Clearly $\mG^+,\mG^-$ have non-zero Poisson brackets with $\mH_\tau,\mH_\sigma$ and hence together form set of second class constraints that vanish strongly. As a result constraints $\mH_\tau=0,\mH_\sigma=0$  can be explicitly solved. We firstly solve $\mH_\sigma=0$ for $\partial_\sigma x^-$ and we get
\begin{eqnarray}\label{partialxminus}
T\partial_\sigma x^-=-p_{y_i}\partial_\sigma y^i-p_{z_m}\partial_\sigma z_m-p_{z_1}\partial_\sigma z_1\equiv -\tmH_\sigma \ . 
\nonumber \\
\end{eqnarray}
Further, Hamiltonian constraint $\mH_\tau=0$ can be solved for $p_+$ which is related to the Hamiltonian on the reduced phase space as $\mH_{red}=-p_+$.
 To do this we insert $\partial_\sigma x^-$ given in (\ref{partialxminus})  into (\ref{mHtaulc}) and using also (\ref{gff}) we get quadratic equation for $p_+$
\begin{eqnarray}
& &-2T(1-a)p_+
\tau^t_{ \ 0}\tau_{z_1}^{ \ 1}\partial_\sigma z_1
+2T^2(1-m_{tz_1}\partial_\sigma z_1)\tau^t_{ \ 0}\tau_{z_1}^{ \ 1}\partial_\sigma z_1-
\nonumber \\
& &-2
(p_{z_1}+m_{z_1t}\tmH_\sigma a)
\tau^{z_1}_{ \ 1}\tau_t^{ \ 0}
(a\tmH_\sigma
)
+\nonumber \\
& &+p_{z_m}(1-\frac{z_1^2}{4R_0^2})^2
p_{z_m}+T^2\partial_\sigma z^m 
\frac{1}{(1-\frac{z_1^2}{4R_0^2})^2}\partial_\sigma z^m+
\nonumber \\
&&+(p_{y_1})^2+T^2(\partial_\sigma y_i)^2
+T^2+2Tp_+a+a^2p_+^2+\tmH_\sigma^2(1-a)^2=0 \nonumber \\
\end{eqnarray}
that can be solved for $p_+$ as
\begin{eqnarray}\label{pplusgen}
p_+=\frac{T(1-a)\tau^t_{ \ 0}\tau_{z_1}^{ \ 1}\partial_\sigma z_1-Ta}{a^2}
-\frac{1}{2a^2}\sqrt{\mK} \ , \nonumber \\
\end{eqnarray}
where we defined $\mK$ as 
\begin{eqnarray}
& &\mK=[2T(1-a)\tau^t_{ \ 0}\tau_{z_1}^{ \ 1}\partial_\sigma z_1-2Ta]^2-4
a^2\left(2T^2(1-m_{tz_1}\partial_\sigma z_1)\tau^t_{ \ 0}\tau_{z_1}^{ \ 1}\partial_\sigma z_1-\right.\nonumber \\
& &-2
(p_{z_1}+m_{z_1t}\tmH_\sigma a)\tau^{z_1}_{ \ 1}\tau_t^{ \ 0}a\tmH_\sigma+
\nonumber \\
& &\left.+p_{z_m}(1-\frac{z_1^2}{4R_0^2})^2p_{z_m}+T^2\partial_\sigma z^m\frac{1}{
(1-\frac{z_1^2}{4R_0^2})^2}\partial_\sigma z^m+
(p_{y_1})^2+T^2(\partial_\sigma y_i)^2+T^2+(1-a)^2\tmH_\sigma^2\right) \ . 
\nonumber\\
\end{eqnarray}
Previous form of the Hamiltonian density on the reduced phase space is valid for $a\neq 0$. Explicitly, it is not valid for $\alpha=-\frac{1}{2}$, that, according to 
\cite{Frolov:2019nrr,Arutyunov:2009ga,Arutyunov:2006gs}, defines temporal gauge  
\begin{equation}
\phi=x^++x^- \ , \quad t=x^+ \ , \quad p_t=p_+-p_- \ , \quad
p_\phi=p_- \ . 
\end{equation}
In this case we should start again with the Hamiltonian constraint (\ref{mHtaulc})
that for $a=0$  has the form 
\begin{eqnarray}
& &\mH_\tau=-2Tp_+
\tau^t_{ \ 0}\tau_{z_1}^{ \ 1}\partial_\sigma z_1
+2T^2(1-m_{tz_1}\partial_\sigma z_1)\tau^t_{ \ 0}\tau_{z_1}^{ \ 1}\partial_\sigma z_1
+\nonumber \\
& &+p_{z_m}(1-\frac{z_1^2}{4R_0^2})^2
p_{z_m}+T^2\partial_\sigma z^m 
\frac{1}{(1-\frac{z_1^2}{4R_0^2})^2}\partial_\sigma z^m+
\nonumber \\
& &+(p_{y_i})^2+(\partial_\sigma y_i)^2
+T^2+(\tmH_\sigma)^2=0 \nonumber \\
\end{eqnarray}
that can be solved for $p_+$ as
\begin{eqnarray}
& &p_+=\frac{1}{2T\tau^t_{ \ 0}\tau_{z_1}^{ \ 1}\partial_\sigma z_1}
[2T^2(1-m_{tz_1}\partial_\sigma z_1)\tau^t_{ \ 0}\tau_{z_1}^{ \ 1}\partial_\sigma z_1+\nonumber \\
& &+p_{z_m}(1-\frac{z_1^2}{4R_0^2})^2p_{z_m}+T^2\partial_\sigma z^m\frac{1}{(1-\frac{z_1^2}{4R_0^2})^2}
	\partial_\sigma z^m
+(p_{y_i})^2+T^2(\partial_\sigma y)^2+T^2+\tmH_\sigma^2] \ . 
\nonumber \\
\end{eqnarray}
Let us return to (\ref{mHtaulc}) and determine its explicit form for some
special cases. For $\alpha=0$ we get uniform light-cone gauge when 
\begin{equation}
\phi=x^++\frac{1}{2}x^- \ , \quad t=x^+-\frac{1}{2}x^- \ , \quad 
p_t=\frac{1}{2}p_+-p_- \ , \quad p_\phi=p_-+\frac{1}{2}p_+ \ , 
\end{equation}
where $p_+$ is equal to
\begin{eqnarray}
& &p_+=2[T\tau^t_{ \ 0}\tau_{z_1}^{ \ 1}\partial_\sigma z_1-T]
-2\sqrt{\mK} \ , \nonumber \\
& &\mK=[T\tau^t_{ \ 0}\tau_{z_1}^{ \ 1}\partial_\sigma z_1-T]^2-
[2T^2(1-m_{tz_1}\partial_\sigma z_1)\tau^t_{ \ 0}\tau_{z_1}^{ \ 1}\partial_\sigma z_1-
(p_{z_1}+\frac{1}{2}m_{z_1t}\tmH_\sigma )\tau^{z_1}_{ \ 1}\tau_t^{ \ 0}\tmH_\sigma+
\nonumber \\
& &+p_{z_m}(1-\frac{z_1^2}{4R_0^2})^2p_{z_m}+T^2\partial_\sigma z^m\frac{1}{
	(1-\frac{z_1^2}{4R_0^2})^2}\partial_\sigma z^m+
(p_{y_1})^2+T^2(\partial_\sigma y_i)^2+T^2+\frac{1}{4}\tmH_\sigma^2] \ . \nonumber\\
\end{eqnarray}
Finally we can consider special case when  $\alpha=\frac{1}{2}$ corresponding
to $a=1$. In this case we find
\begin{eqnarray}
& &p_+=-T-
\frac{1}{2}\sqrt{\mK} \ , \nonumber \\
& &\mK=4T^2-4
(2T^2(1-m_{tz_1}\partial_\sigma z_1)\tau^t_{ \ 0}\tau_{z_1}^{ \ 1}\partial_\sigma z_1-2
(p_{z_1}+m_{z_1t}\tmH_\sigma )\tau^{z_1}_{ \ 1}\tau_t^{ \ 0}a\tmH_\sigma+
\nonumber \\
& &+p_{z_m}(1-\frac{z_1^2}{4R_0^2})^2p_{z_m}+T^2\partial_\sigma z^m\frac{1}{
	(1-\frac{z_1^2}{4R_0^2})^2}\partial_\sigma z^m+
(p_{y_1})^2+T^2(\partial_\sigma y_i)^2+T^2) \ . 
\nonumber\\
\end{eqnarray}
We see that the case $\alpha=-\frac{1}{2}$ is exceptional since in this case
the Hamiltonian density on the reduced phase space is quadratic in momenta while
generally the Hamiltonian has square root structure. In the next section we will
analyse some classical solutions of the equations of motion on the reduced phase space. 

\section{Properties of Non-Relativistic String in Uniform Light-Cone gauge}\label{fourth}
In this section we will discuss properties of  non-relativistic string theory 
on reduced phase space.
First of all we consider equations of motion for $y_i,p_{y_i},z_m,p_m$ that due to the fact that the background fields do not depend on them have the form
\begin{eqnarray}
& &\partial_\tau y_i=\pb{y_i,H_{red}}=\frac{1}{2a^2\sqrt{\mK}}
(T^2p_{y_i}+(\dots) \partial_\sigma y_i) \ , \nonumber \\
& &\partial_\tau z_m=\pb{z_m,H_{red}}=\frac{1}{2a^2\sqrt{\mK}}
((1-\frac{z_1^2}{4R_0^2})^2p_{z_m}+(\dots)\partial_\sigma z_m) \ , 
\nonumber \\
\end{eqnarray}
where 
$H_{red}=\int d\sigma \mH_{red}$ and where $(\dots)$ mean terms which are not important for us.  The equations above can be solved by the ansatz
 $z_m=p_{z_m}=y_i=p_{y_i}=0$. In fact, this ansatz also solves equations of motion for $p_{z_m}$ and $p_{y_i}$.
As a result we can consider Hamiltonian density for $z_1$ only that has the form
\begin{eqnarray}\label{mHredz1}
& &\mH_{red}^{z_1}=-\frac{T(1-a)\tau^t_{ \ 0}\tau_{z_1}^{ \ 1}\partial_\sigma z_1-Ta}{a^2}+\frac{1}{2a^2}\sqrt{\mK} \ , \nonumber \\
& &\mK=[2T(1-a)\tau^t_{ \ 0}\tau_{z_1}^{ \ 1}\partial_\sigma z_1-2Ta]^2-\nonumber \\
& &-4
a^2(2T^2\tau^t_{ \ 0}\tau_{z_1}^{ \ 1}\partial_\sigma z_1-2
p_{z_1}\tau^{z_1}_{ \ 1}\tau_t^{ \ 0}a p_{z_1}\partial_\sigma z_1
+T^2+(1-a)^2(p_{z_1}\partial_\sigma z_1)^2) \ . 
\nonumber\\
\end{eqnarray}
In order to find equation of motion for $z_1$ it is convenient to find Lagrangian from (\ref{mHredz1}). To do this we firstly determine  canonical  equation of motion using (\ref{mHredz1})
\begin{eqnarray}
\partial_\tau z_1=\frac{1}{4a^2\sqrt{\mK}}
(8a^2\tau^{z_1}_{ \ 1}\tau_t^{ \ 0}
a\partial_\sigma z_1-2(1-a)^2(\partial_\sigma z_1)^2)p_{z_1} \ .
\nonumber \\
\end{eqnarray}
Then $\mL_{red}^{z_1}$ is given by standard formula
\begin{eqnarray}\label{mLred}
& &\mL_{red}^{z_1}=p_{z_1}\partial_\tau z_1-\mH^{z_1}_{red}
=\frac{T(1-a)\tau^t_{ \ 0}\tau_{z_1}^{ \ 1}\partial_\sigma z_1-Ta}{a^2}-\nonumber \\
& &
-\frac{T}{a}\sqrt{(1-a)^2(\tau^t_{ \ 0}\tau_{z_1}^{ \ 1}\partial_\sigma z_1)^2-
	2a\tau^t_{ \ 0}\tau_{z_1}^{ \ 1}\partial_\sigma z_1+a^4(\tau_{z_1}^{ \ 1}\tau^t_{ \ 0})^2(\partial_\tau z_1)^2}=
\nonumber \\
&&=\frac{T(1-a)g\partial_\sigma z_1-Ta}{a^2}-\frac{T}{a}\sqrt{\bB} \ , 
\nonumber \\
\end{eqnarray}
where we introduced $g$ and $\bB$ defined as
\begin{equation}
g\equiv \tau^t_{ \ 0}\tau_{z_1}^{ \ 1} \ , \quad 
\bB=(1-a)^2g^2(\partial_\sigma z_1)^2-2ag\partial_\sigma z_1+a^2g^2(\partial_\tau z_1)^2 \ . 
\end{equation}
Note also that (\ref{mLred}) is valid for $a\neq 0$. 
Then performing variation of (\ref{mLred}) we get following equation of motion for 
$z_1$
%
\begin{eqnarray}
& &-\frac{1}{\sqrt{\bB}}(1-a)^2gg'(\partial_\sigma z_1)^2
+\partial_\sigma[\frac{1}{\sqrt{\bB}}g^2\partial_\sigma z_1]-\frac{ag}{2}\frac{1}{\bB^{3/2}}\partial_\sigma \bB-\nonumber \\
& &-a^4\frac{gg'}{\sqrt{\bB}}(\partial_\tau z_1)^2+
a^4\partial_\tau[\frac{g^2}{\sqrt{\bB}}\partial_\tau z_1]=0 \ . 
\nonumber \\
\end{eqnarray}
This is rather complicated equation and it is difficult to solve it in the full
generality. On the other hand we can certainly gain  in sign into its form when
we consider some simpler ansatz as for example $z_z=z_z(\tau)$. However it turns out that this is too restrictive since it is easy to see that the equation above is solved for any $z_1(\tau)$. It is possible that more general ansatz could be desirable but we are not going proceed along this way. 

we rather proceed to the  more interesting situation when 
$a=0$. It is easy to see that as in general case $a\neq 0$ we can consistently 
set
 $p_m=z_m=y_i=p_{y_i}=0$ so that the reduced Hamiltonian density has the form
\begin{eqnarray}
\mH_{a=0}^{z_1}=-\frac{1}{2Tg\partial_\sigma z_1}
[2T^2g\partial_\sigma z_1
+T^2+p_{z_1}^2(\partial_\sigma z_1)^2] \ .
\nonumber \\
\end{eqnarray}
Then  the first canonical equation has the form
\begin{equation}
\partial_\tau z_1=\pb{z_1,H_{a=0}^{z_1}}=
-\frac{1}{Tg}p_{z_1}\partial_\sigma z_1
\end{equation}
so that Lagrangian density has the form 
\begin{eqnarray}
\mL_{a=0}^{z_1}=p_{z_1}\partial_\tau z_1-\mH_{a=0}^{z_1}=
-\frac{Tg}{2\partial_\sigma z_1}
(\partial_\tau z_1)^2
+T\partial_\sigma z_1 \ .  \nonumber \\
\end{eqnarray}
It is easy to derive corresponding equation of motion for $z_1$
\begin{equation}
-\frac{T}{2}\frac{dg}{dz_1}\frac{(\partial_\tau z_1)^2}{\partial_\sigma z_1}
-T\partial_\sigma\left[\frac{g}{2}\frac{(\partial_\tau z_1)^2}{(\partial_\sigma z_1)^2}\right]+T\partial_\tau\left[\frac{g}{\partial_\sigma z_1}\partial_\tau z_1\right]=0 \ . 
\nonumber \\
\end{equation}
Clearly this equation has solution $z_1=z_1(\sigma)$ for any function $z_1$. On the other hand let us consider an ansatz $z_1=f(\sigma-v\tau)$ so that $\partial_\tau z_1=-vf' \ , \quad \partial_\sigma z_1=f'$ and hence the equation of motion has the form
\begin{eqnarray}
-\frac{T}{2}\frac{dg}{dz_1}v^2 f'-\frac{T}{2}v^2\partial_\sigma g-
Tv\partial_\tau g=
-T\frac{dg}{dz_1}v^2f'+Tv^2\frac{dg}{dz_1}f'=0 \ . 
\nonumber \\
\end{eqnarray} 
In other words this equation is obeyed by any function $f(\sigma-v\tau)$. This is again interesting property of non-relativistic
string in uniform light-cone gauge.

\section{Lagrangian Equations of Motion}\label{fifth}
In this section we find equations of motion of new-non-relativistic string with application to the $AdS_5\times S^5$ case. Recall that the Lagrangian has the form 
\begin{equation}\label{SnonfinalL}
S=-\frac{T}{2}\int d^2\sigma [\sqrt{-\tau}\tau^{\alpha\beta}h_{\alpha\beta}+
\epsilon^{\alpha\beta}m_{\alpha\beta}]  \ .
\end{equation}
From this action we derive following equations of motion for $x^M$ 
\begin{eqnarray}
& &\frac{1}{2}\sqrt{-\tau}
\partial_M \tau_{KL}\partial_\alpha x^K\partial_\beta x^L\tau^{\beta\alpha}
\tau^{\gamma\delta}h_{\gamma\delta}-\partial_\alpha[\sqrt{-\tau}\tau^{\alpha\beta}\tau_{MN}\partial_\beta x^N
\tau^{\gamma\delta}h_{\gamma\delta}]+
\nonumber \\
& &+\sqrt{-\tau}\tau^{\alpha\beta}\partial_M h_{KL}\partial_\alpha x^K
\partial_\beta x^L+\epsilon^{\gamma\delta}\partial_M m_{KL}\partial_\gamma x^K
\partial_\delta x^L-\nonumber \\
& &-\sqrt{-\tau}\tau^{\alpha\delta}\partial_M \tau_{KL}\partial_\delta x^K
\partial_\gamma x^L\tau^{\delta \beta}h_{\beta\alpha}+
2\partial_\alpha[\sqrt{-\tau}\tau^{\gamma\alpha}\tau_{MN}\partial_\beta x^N
\tau^{\beta \delta}h_{\delta\gamma}]-\nonumber \\
& &-2\partial_\alpha[\sqrt{-\tau}\tau^{\alpha\beta}h_{MN}\partial_\beta x^N]
-2\partial_\alpha[\epsilon^{\alpha\beta}m_{MN}\partial_\beta x^N]=0 \ . 
\nonumber \\
\end{eqnarray}
These equations of motion are very complicated in the full generality 
and we rather proceed in different way when we try to analyse non-relativistic string in $AdS_5\times S^5$ background. As in previous section we consider 
an ansatz   $z_m=y_i=0$ so that we are interested in dynamics of $t$ and $z_1$ only. We further presume that world-sheet time $\tau$ coincides with $t$ and also that $\phi$ coincides with $\sigma$. Explicitly, we presume following ansatz 
\begin{equation}
t=\kappa \tau \ , \quad \phi=\sigma \ , \quad z_1=f(\sigma-v\tau) \ . 
\end{equation}
Inserting this ansatz into action (\ref{SnonfinalL}) we obtain
\begin{equation}\label{SnonfinalLin}
S=\frac{T}{\kappa}\int d^2\sigma 
\left[g\frac{1}{f'}+\frac{v^2}{g}f'\right]  \ , \quad g=\sqrt{-\frac{\tau_{tt}}{\tau_{z_1z_1}}} \ , \quad f'\equiv \frac{df}{dx} \ , \quad x=\sigma-v\tau \ . 
\end{equation}
Performing variation of (\ref{SnonfinalLin}) with respect to $f$ we get following
equation
\begin{eqnarray}
\frac{dg}{df}\frac{1}{f'}+\frac{d}{dx}\left(\frac{g}{f'^2}\right)
-\frac{v^2}{g^2}\frac{dg}{df}f'-v^2\frac{d}{dx}\left(\frac{1}{g}\right)
=0\nonumber \\
\end{eqnarray}
that can be simplified into the form 
\begin{eqnarray}
\frac{1}{f'}\frac{d}{dx}\left(\frac{g}{f'}\right)=0 \ . 
\nonumber \\
\end{eqnarray}
The first integral of this equation is equal to
\begin{equation}
\frac{g}{f'}=K \ ,  
\end{equation}
where $K$ is constant. Then using the fact that  $g(f)=1+\frac{f^2}{4R_0^2}$
we obtain final result 
\begin{equation}
	f=4R_0|\tan \frac{K(\sigma-v\tau)}{4R_0}| \ . 
\end{equation}
Note that this is similar solution as was discussed recently in \cite{Fontanella:2021btt} and that has physical interpretation as an infinite
array of spikes on the world-sheet of non-relativistic string that is extended
along $\phi$ direction. Certainly it could be possible to consider
more general ansatz and analyse corresponding equations of motion  in the similar
way as in \cite{Fontanella:2021btt}.
\\
\\
{\bf \Large Appendix: Non-relativistic string in $AdS_5\times S^5$ in an alternative set of coordinates}\label{sixth}
\\
\\
In this appendix we present formulation of non-relativistic string
in $AdS_5\times S^5$ background using coordinates that were introduced in  \cite{Gomis:2005pg}. In this formulation the  vector indices for $AdS_5$ are labelled with $m=0,1,2,3,4$ while for
$S^5$ we have $m'=1',2',3',4',5'$ with the flat metric
$\eta_{mn}=\mathrm{diag}(-1,1,1,1,1)$ and $\delta_{m'n'}=\mathrm{diag}(1,1,1,1,1)$. Then vielbein has following form
\begin{eqnarray}
& &e^{ \ 0}=dT\cosh \rho \ , \quad  \rho=\frac{\sqrt{X^a\eta_{ab}X^b}}{R} \ , \quad 
e^{ \ 1}=dX^1\cosh \rho \cos \frac{T}{R} \ , \nonumber \\
& &e^{ \ a}=dX^a+dX^b(\eta_b^{ \ a}-\frac{X_b X^a}{\rho^2 R^2})
(\frac{\sinh \rho}{\rho}-1) \  , \nonumber \\
& &e^{ \ m'}=dX^{m'}+dX^{n'}(\eta_{n'}^{ \ m'}-\frac{X_{n'}X^{m'}}{r^2R^2})
(\frac{\sin r}{r}-1) \ , \quad r=\frac{\sqrt{X^{m'}\delta_{m'n'}X^{n'}}}{R} \ ,
\nonumber \\
\end{eqnarray}
where $a=2,3,4$. 
Let us now take non-relativistic limit in the form
\begin{equation}
T=\omega t \ , \quad  X^1=\omega x^1 \ ,   \quad   R=\omega R_0 \ ,
\end{equation}
so that we obtain
\begin{eqnarray}
e^{ \ 0}=cdt (1+\frac{\hat{\rho}^2}{2c^2}) \ , 
\nonumber \\
\end{eqnarray}
from which we can deduce following components of $\tau_M^{ \ A}$ and $\pi_M^{ \ A}$
\begin{equation}\label{tauG1}
\tau_t^{\ 0}=1 
\ , \quad \pi_t^{\ 1}=-\hat{\rho}^2 \ , \quad \hat{\rho}=\sqrt{x^a x_a}/R_0 \ . 
\end{equation}
We further have
\begin{equation}
e^1=cdx^1(1+\frac{\hat{\rho}^2}{2c^2})
\cos \frac{t}{R_0}
\end{equation}
that again implies
\begin{equation}\label{tauG2}
\tau_1^{ \ 1}=\cos \frac{t}{R_0} \ , \quad \pi_1^{ \ 0}=-\hat{\rho}^2\cos\frac{t}{R_0}
\ .  
\end{equation}
Finally we have
\begin{equation}
e^{ \ a}=dx^a+dx^b(\eta_b^{  \ a}-\frac{x_b x^a}{\hat{\rho}^2R_0^2})
(1+O(c^{-3})-1)=dx^a \ .
\end{equation}
Now we are ready to proceed to find corresponding Hamiltonian.
As the first step we determine following non-zero  components of $m_{MN}$ 
\begin{eqnarray}
m_{t1}=
\hat{\rho}^2\cos\frac{t}{R_0} \  .
\end{eqnarray}
Further, from the relation 
$\tau_M^{ \ A}\tau^M_{ \ B}=\delta^A_B$ we obtain that there are non-zero components
\begin{eqnarray}
\tau^t_{ \ 0}=\frac{1}{\tau_t^{ \ 0}}=1 \ , \quad
\tau^{1}_{ \ 1}=\frac{1}{\cos\frac{t}{R_0}}
\nonumber \\
\end{eqnarray}
so that Hamiltonian constraint is equal to
\begin{eqnarray}
&&\mH_\tau=-2T\Pi_t\cos\frac{t}{R_0}\partial_\sigma x^1-2T\Pi_1 \frac{1}{\cos\frac{t}{R_0}}\partial_\sigma t+
\nonumber \\
&&+p_{x^a}\delta^{ab}p_{x^b}+T^2\partial_\sigma x^a\delta_{ab}\partial_\sigma x^b+
p_{m'}\delta^{m'n'}p_{n'}+T^2\partial_\sigma x^{m'}\delta_{m'n'}\partial_\sigma x^{n'} \ , 
\nonumber \\
\end{eqnarray}
where
\begin{equation}
\Pi_t=p_t+Tm_{t1}\partial_\sigma x^1 \ , \quad 
\Pi_{1}=p_{1}+Tm_{1 t}\partial_\sigma t \ . 
\end{equation}
Form of the  Hamiltonian constraint  implies that it is not possible 
to impose uniform light-cone gauge due to its explicit dependence on $t$. Certainly it is possible to study non-relativistic string in the background defined 
by (\ref{tauG1}) and (\ref{tauG2}) but the Lagrangian formulation is the same
as in \cite{Gomis:2005pg}  which is well known and hence we will not study it in this paper.

{\bf Acknowledgement:}
\\
This work 
is supported by the grant “Integrable Deformations”
(GA20-04800S) from the Czech Science Foundation
(GACR).

\end{document}